\documentclass[aps,twocolumn]{revtex4-1}
\usepackage{graphicx}

\begin{document}

\title{Topology Sorting and Characterization of Folded Polymers Using Nano-pores}

\author{Narges Nikoofard}
\affiliation{Institute of Nanoscience and Nanotechnology, University of Kashan, Kashan 51167-87317, Iran.}

\author{Alireza Mashaghi}
\email{Alireza\_Mashaghi-Tabari@meei.harvard.edu}
\affiliation{Harvard Medical School, Harvard University, Boston, MA 02115, USA.}
\date{\today}

\begin{abstract}

Here we report on the translocation of folded polymers through nano-pores using molecular dynamic simulations. Two cases are studied; one in which a folded molecule unfolds upon passage and one in which the folding remains intact as the molecule passes through the nano-pore. The topology of a folded polymer chain is defined as the arrangement of the intramolecular contacts, known as circuit topology. In the case where intramolecular contacts remain intact, we show that the dynamics of passage through a nano-pore varies for molecules with differing topologies: a phenomenon that can be exploited to enrich certain topologies in mixtures. We find that the nano-pore allows reading of topology for short chains. Moreover, when the passage is coupled with unfolding, the nano-pore enables discrimination between pure states, i.e., states for which the majority of contacts are arranged identically. In this case, as we show here, it is also possible to read the positions of the contact sites along a chain. Our results demonstrate the applicability of nano-pore technology to characterize and sort molecules based on their topology.

\end{abstract}

\maketitle

\section{Introduction}

Most biopolymers, such as RNAs  \cite{rna}, proteins \cite{protein,protein2} and genomic DNA \cite{genome}, are found in folded configurations. Folding involves the formation of one or more intramolecular interactions, termed contacts. Proper folding of these molecules is often necessary for their function. Intensive efforts have been made to measure the geometric and topological properties of protein and RNA folds, and to find generic relations between those properties and molecular function, dynamics and evolution \cite{rna,protein,protein2,genome}. Likewise, topological properties of synthetic molecules have been subject to intense research, and their significance for polymer chemistry \cite{synthetic,synthetic2} and physics \cite{physics1,physics2,physics3} has been widely recognized.   
 
\begin{figure*}
\centering
\includegraphics[scale=1]{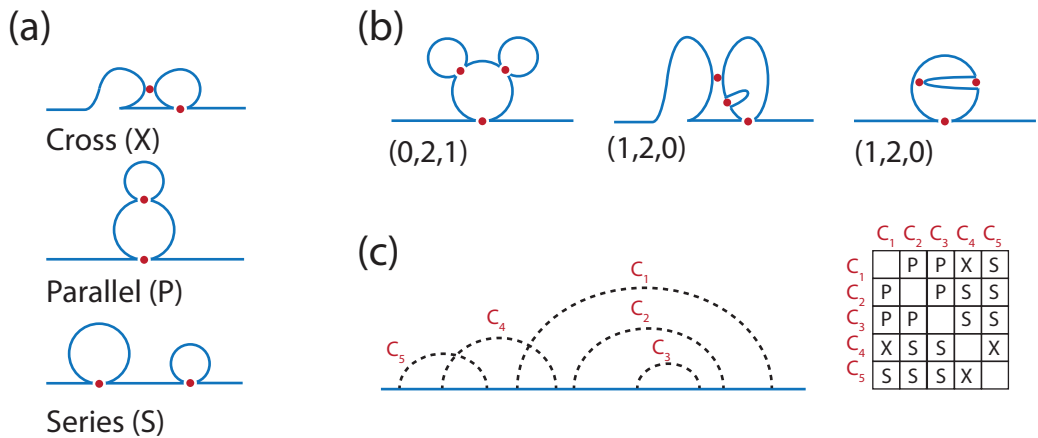}
\caption{(a) Three topologically different configurations of a 2-contact chain. (b) Three selected configurations of a 3-contact chain that have two parallels out of three total topology relations. The numbers in parenthesis are the frequencies of cross, parallel and series topologies ($N_c$,$N_p$,$N_s$), respectively.  (c) Topology of any given n-contact chain can be defined by assigning one of the three binary relations described in panel (a) to each contact pair. For a chain with $N$ contacts, one can identify $\frac{N(N-1)}{2}$ binary relations. Here, the dashed lines correspond to the sites to be connected along the chain (solid line). The topology matrix of the chain is shown on the right.} \label{fig1}
\end{figure*}

Topology is a mathematical term, which is used to describe the properties of objects that remain unchanged under continuous deformation \cite{topology}. Different approaches have been discussed in the literature to describe the topology of branched \cite{branch} or knotted polymers \cite{knot}. However, many important biopolymers, such as proteins and nucleic acids, are unknotted linear chains. The circuit topology approach has recently been introduced to characterize the folded configuration of linear polymers. Circuit topology of a linear chain elucidates generically the arrangement of intra-chain contacts of a folded-chain configuration \cite{mashaghi2014} (see Fig. \ref{fig1}). 
The arrangement of the contacts has been shown to be a determinant of the folding rates and unfolding pathways of biomolecules, \cite{mugler2014} and has important implications for bimolecular evolution and molecular engineering \cite{mashaghiRSC2015, mashaghiSM2015}.   
 
Topology characterization and sorting of polymers has been the subject of intense research in recent years; bulk purification of theta-shaped and three-armed star polymers is performed using chromatography \cite{AnalChem2008,Polymer2009}; linear and circular DNA are separated in nano-grooves embedded in a nano-slit \cite{nl2011}; and star-branched polymers with different number of arms are shown to travel with different speeds through a nano-channel \cite{ma2010}. In the context of characterization, linear and circular DNA molecules are probed by confining them in a nano-channel and using fluorescence microscopy \cite{nl2009}. We know little about how to sort folded linear polymers based on topology. This is in contrast to size sorting of folded linear polymers which has been studied extensively in the literature \cite{PNAS2010,ChemSocRev2010,ChemRev2013}.
 
Nano-pore technology represents a versatile tool for single-molecule studies and biosensing. A typical setting involves a voltage difference across the nano-pore in an ionic solution containing the desired molecule. The ion current through the nano-pore decreases as the molecule enters the pore. The level of current reduction and its duration reveals information about the molecule \cite{nano-pore,solid}.  
Prior to the current project, different properties of nucleic acids and proteins have been studied using nano-pore technology, for example: DNA sequencing \cite{sequencing,sequencing2}, unzipping of nucleic acids \cite{unzipping,unzipping2}, protein detection \cite{detection,detection2}, unfolding of proteins \cite{unfolding1,unfolding2,unfolding3}, and interactions between nucleic acids and proteins \cite{nucleosome,TF}. 

In our study, we used simple models of polymer chains and molecular dynamic simulations to determine how the circuit topology of a chain influences its passage through a nano-pore. We investigated whether nano-pores can be used for topology-based sorting and characterization of folded chains. Two scenarios were considered: (1) passage through pores large enough to permit the chain to pass through without breaking its contacts, and (2) passage of chains through small nano-pores, during which contacts were ripped apart. In the first scenario, nano-pore technology enabled purification of chains with certain topologies and allowed us to read the topology of a folded molecule as it passed through the pore. In the second scenario, we used the nano-pore to read the circuit topology of a single fold. We also asked if translocation time and chain topology are correlated. This technology has been subject to intense research for simple-structured polynucleotides \cite{gerland2012}; however, the current study is the first to use nano-pores to systematically measure contact arrangements of folded molecules \cite{mashaghi2014} (Fig. \ref{fig1}).  

\section{Simulation method}
The polymer is modeled by beads connected by FENE bonds $U_{FENE}=-\frac{1}{2}KR_0^2\ln\left[1-\left(\frac{r}{R_0}\right)^2\right]$. $K$ and $R_0$ are the strength and the maximum extension of the bonds, respectively.  The FENE potential is used to eliminate unrealistic extension of the bonds due to the pulling event. The short-range repulsive interaction between monomers is taken into account by the shifted-truncated Lennard-Jones potential $U_{LJ}=4\epsilon_0\left[\left(\frac{b}{r}\right)^{12}-\left(\frac{b}{r}\right)^{6}+\frac{1}{4}\right]$ at $r<2^{\frac{1}{6}}b$. $\epsilon_0=k_BT$ is the energy scale of the simulations. $b$ is the monomer size and the length scale of the simulations. 

All simulations were performed by ESPResSo \cite{espresso} as detailed below. Initially, the first monomer is fixed inside the nano-pore. After the whole polymer is equilibrated, the first monomer is unfixed and force, $F$, is applied to pull it through the nano-pore. For pore diameters smaller than two monomers, passage of the polymer inevitably leads to breakage of the contacts. In this case, the bond between the contact sites is replaced with a simple Lennard-Jones potential $U_{LJ}=4\epsilon\left[\left(\frac{b}{r}\right)^{12}-\left(\frac{b}{r}\right)^{6}\right]$ after equilibration. The depth of the attraction well, $\epsilon$, is a measure of the strength of the bond between the contact sites.

Number of passed monomers and position of the first monomer versus time are studied in simulations. These quantities are averaged over different realizations. For longer passages, the averages are again window-averaged over intervals equal to 10 time units. Window-averaging is used to reduce the data points and the noise in the plots. To minimize the effect of determinants other than topology, we take equal spacing, $l$, between the contact sites (connected monomers) and two tails on the sides equal to the spacings. The total length of the polymer is $5l+4$. If the monomers are numbered consecutively from one end, then, the position of the contact sites along the chain would be $l+1$, $2l+2$, $3l+3$ and $4l+4$. The spacing is taken equal to 12 monomers, unless otherwise stated. Some chains become knotted when the bonds are formed in the chain or when the chain is pulled suddenly with a strong force.  Passage times of these knotted chains are extremely long. The data related to these unusual passages is removed before averaging.

\section{Results and discussions}
\subsection{Passage of folded chains}

\begin{figure}
\centering
\includegraphics[scale=0.8]{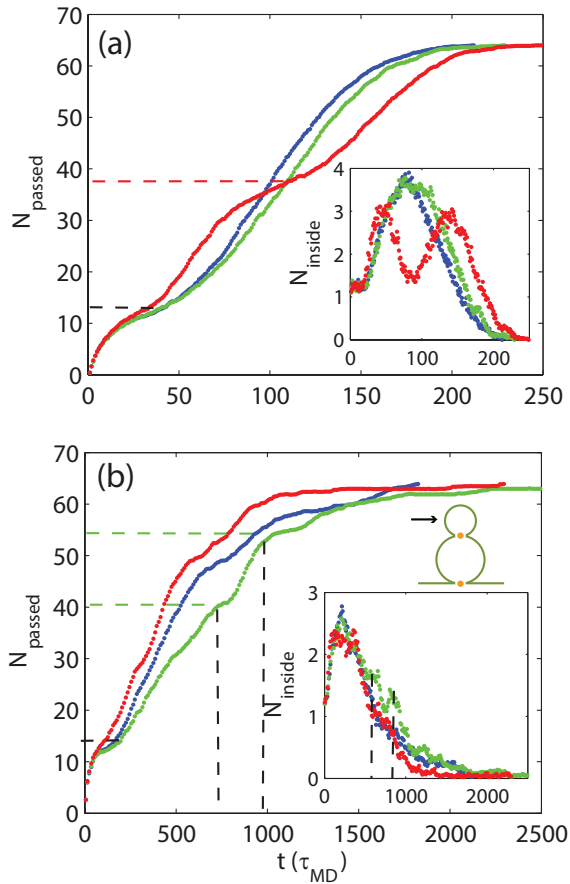}
\caption{Average number of monomers passed through the nano-pore versus time for a chain with two unbreakable contacts. The pulling forces are equal to (a) $F = 10\frac{k_BT}{b}$ and (b) $F = 6\frac{k_BT}{b}$. The shoulders correspond to pauses in the passage process and can be used to read the chain topology. The series topology shows two shoulders under the strong force (a) and the parallel topology shows three shoulders under the weak force (b). Insets: Average number of monomers inside the nano-pore versus time. The peaks in the inset plots occur at the same time with the shoulders in the main plots. Schematics shows the parallel topology and the arrow points to the smaller loop. Entropy of the smaller loop causes the last pause in the passage of the parallel topology under the weak force.} \label{fig2}
\end{figure}

Consider translocation of 2-contact chains through a nano-pore with an internal diameter equal to $3b$. First we assume contacts are permanent and unbreakable. Two different strengths for the pulling force, $F = 6\frac{k_BT}{b}$ and $10\frac{k_BT}{b}$, are examined. 
50 realizations are performed for each of the three topologies (Fig. \ref{fig1}(a)) and the two forces. The average number of monomers passed through the nano-pore versus time is shown in Figs. \ref{fig2}(a) and \ref{fig2}(b). Shoulders in the curves correspond to pauses during the passage of the polymer when the contacts encounter the nano-pore. We first examine the passage dynamics under stronger force, $F = 10\frac{k_BT}{b}$ (Fig. \ref{fig2}(a)). One shoulder is observed during the passages of the cross and the parallel topologies, while the series topology is markedly different with two clear shoulders during its passage. The average number of passed monomers at the shoulders coincides with the position of the contact sites (shown with horizontal lines in the plot). This confirms interpretation of the shoulders as the pauses related to the passage of the contacts. The average number of monomers inside the nano-pore versus time is also significantly different for the series topology (inset of Fig. \ref{fig2}(a)). Two distinct peaks are observed for the series topology, while only one peak is seen for the cross and the parallel topologies. The peaks in the inset plot occur simultaneously with the shoulders in the main plot.

Force has a dramatic effect on the passage dynamics of chains. Fig. \ref{fig2}(b) plots the average number of monomers passed through the nano-pore versus time under $F = 6\frac{k_BT}{b}$. Here, the maximum passage time for the chain with parallel topology is larger relative to other topologies, while the one with the series topology had the largest maximum passage time under the stronger force. By reducing the force, the passage time gets much longer and the entropic effects become dominant. As a result, the shoulders in the number of passed monomers are not as clear as before. One shoulder is observed for all topologies at the position of the first contact site. Two other shoulders are observed during the passage of the chain with parallel contact arrangemenet. The second shoulder corresponds to the time when the large loop of the chain is midway inside the nano-pore. Furthermore, the third shoulder is due to the second contact and the pause caused by the entropy of the second small loop in the parallel topology (shown schematically in the inset). Additionally, two peaks are seen in the time profile of the average number of monomers inside the nano-pore. These peaks appear simultaneously with the described shoulders in the time profile of the number of passed monomers (inset of Fig. \ref{fig2}(b)).
We note that the first and the third shoulders are also seen in the average position of the first monomer versus time (Fig. S2).

The results show that it is possible to distinguish the series and parallel from other topologies using nano-pores with strong and weak forces, respectively. The number of passed monomers and the number of monomers inside the nano-pore can be readily measured in experiments. The former can be measured by pulling the end of the polymer using optical tweezers \cite{solid}, while the latter is readable by measuring the ion current through the nano-pore. The ion current has been shown to take discrete values with the number of monomers in the nano-pore \cite{nano-microscope}. Finally, the maximum passage time in each pulling force is also different for the three topologies and can be used alternatively for identifying the topology of a 2-contact unbreakable chain. To generalize the obtained results to molecules of various sizes (chain lengths), we investigate the passage of a chain with two unbreakable contacts and double spacing between the contact sites under weak and strong pulling forces. 
Position of the shoulders and the peaks are in agreement with the above descriptions (Fig. S3). Also, it is seen that the maximum passage time is longer for the series topology under the strong force and for the parallel topology under the weak force. (see ESI section 1)

Next, we consider folded molecules with more than two intramolecular contacts. Nano-pores of various sizes are needed to pass these complex unbreakable topologies under usual pulling forces. This gives the opportunity to use the nano-pore for purifying topologies or for enrichment of a certain topology from a mixture of different topologies. To test this idea, we examine the passage of 3-contact chains through a pore with an internal diameter equal to $3b$ under the pulling force $F = 10\frac{k_BT}{b}$. There are 15 topologically different configurations for a 3-contact chain. Among these, three configurations have two parallel relations in their topologies, shown in Fig 1(b). Two of them (among all 15 configurations) do not pass the nano-pore in usual time intervals. This is in agreement with the expectation that chains with a higher number of parallel topologies tend to interlock more, and do not pass through smaller pores. The three chains shown in Fig. \ref{fig1}(b) behave similarly when they enter the nano-pore from either end. This means that the chain direction is not important in purifying the topologies using a nano-pore.

\begin{figure}
\centering
\includegraphics[scale=0.8]{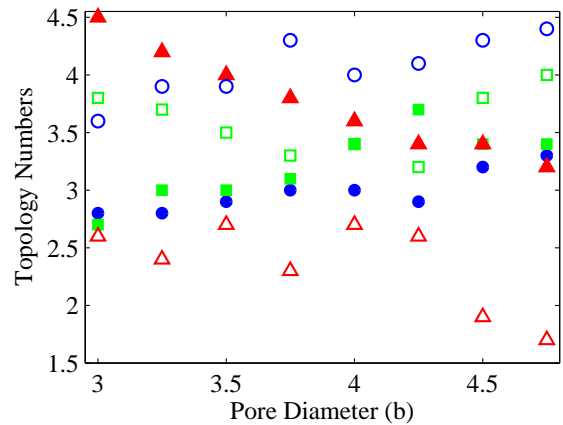}
\caption{Average number of topologies in the chains versus the internal diameter of the nano-pore. Filled and empty symbols correspond to the chains that pass and do not pass through the nano-pore, respectively. The average number of cross, parallel and series topologies are shown with circles, squares and triangles, respectively. The pulling force is equal to  $F = 10\frac{k_BT}{b}$.  For the smaller nano-pores, 150 random configurations are averaged. However, up to 700 random configurations are tested for the nano-pores with $d=4.5b,4.75b$. All the chains can pass through the nano-pore when the nano-pore diameter is larger than or equal to $5b$.  } \label{fig3}
\end{figure}

We then extended the simulations to 5-contact chains as a representation of real chains with increasing complexities. There are (2*5-1)!!=1500 topologically different configurations for a 5-contact chain, so we chose configurations at random and passed them through the nano-pore under the pulling force $F = 10\frac{k_BT}{b}$. The chains are examined to see whether they pass through the nano-pore in a reasonable time. 
A first-order measure of the circuit topology of a chain is the number of contact pairs that are in series, cross, or parallel arrangements. Fig. 3 shows the average numbers of the three topological arrangements versus the internal diameter of the nano-pore, $d$, for chains that pass and do not pass through the nano-pore. The average number of series topology is higher in the passed chains compared to the chains that do not pass. The average number of parallel topology is smaller in the passed chains, for pore diameters smaller than four times the monomer size. Average number of cross topology is smaller for passed chains. In a realistic setting, when a mixture of randomly connected chains are allowed to pass through a nano-pore (smaller than $4b$), we predict that the flow through would be enriched in series topology. However, the fraction of the mixture that fails to pass through the pore would contain chains with high number of parallel and cross topologies. 

This can be justified by the fact that in parallel and cross topologies, the contact sites are relatively far from each other along the chain. Thus, chains with a high number of contacts with parallel and cross arrangements are bulkier and have more interlocking configurations. In contrast, in series topology, the contacts are local and do not connect distant points along the chain. As a result, the chains with a high number of contacts with series arrangements are more extended in configuration and can pass through the nano-pore more easily. As it is evident from our study, the excluded volume interaction is the main deriving force behind separation in a narrow nano-pore; this interaction has not been considered in previous theoretical works \cite{AnalChem2008,Polymer2009}.

Finally, simulations with a four times stronger pulling force $F = 40\frac{k_BT}{b}$ shows that no purification is possible under higher forces even with the smallest nano-pore, $d=3b$, reflecting the importance of tuning the applied force to its optimal values.

\subsection{Passage coupled to chain unfolding}

There are two parameters that determine the time needed for the passage of a chain coupled to bond breakup; the bond strength and the pulling force.
We first studied pulling of 2-contact chains through a nano-pore with internal diameter equal to $1.4b$, under a force comparable to the bond strength  (see ESI section 2 for a theoretical description). 
For very weak bonds, with a bond strength equal to $5k_BT$, the contacts break before reaching the pore. This is due to the tension propagated along the chain from the pulled end \cite{sakaue2007}. For medium to strong bonds between $10-40k_BT$, it is possible to see shoulders in the time profile of the position of the first monomer, using suitable pulling forces (Figs. 4(a) and 4(b)). For shoulders to become prominent, a large pulling force is required to dominate the entropic fluctuations. However, it should not be too strong to completely eliminate the effect of topology. As the leading end of the chain is stretched completely with large pulling forces, the shoulders can be used to find position of the contact sites along the chain (horizontal lines in Figs. 4(a) and 4(b)).

\begin{figure}
\centering
\includegraphics[scale=0.8]{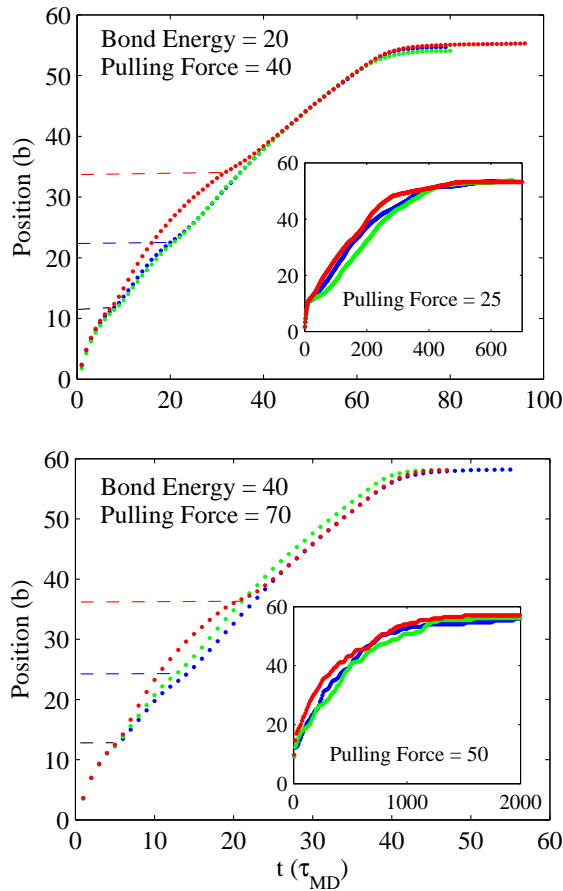}
\caption{Average position of the first monomer versus time, for the bond energies $20k_BT$ and $40k_BT$. The pulling force should be chosen carefully: large enough to minimize the effect of entropy  but not too large to eliminate the effect of topology. The shoulders are due to the pauses at the contact sites. Position of the shoulders can give information about the position of the connected monomers along the chain (horizontal lines). Insets: Position of the contact sites cannot be tracked in smaller pulling forces.} \label{fig4}
\end{figure}

\begin{figure}
\centering
\includegraphics[scale=0.8]{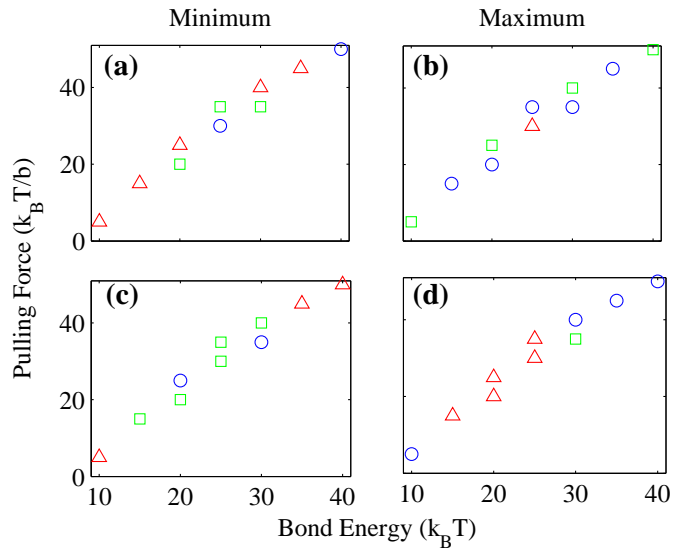}
\caption{(a) and (b) Topologies that have the minimum and the maximum passage times in the translocation of 2-contact chains with different topologies. Passage times are averaged over 50 realizations. (c) and (d) Dominant topologies with the minimum and the maximum passage times in the translocation of 5-contact chains with one dominant topology. The dominant topology comprises 8 out of 10 total binary relations in the chain. The passage times are averaged over 150 random configurations with the same dominant topology. Circle, square and triangle symbols correspond to cross, parallel and series topologies, respectively. } \label{fig5}
\end{figure}

For large forces, there is no difference between the passage times of the three topologies. The difference increases by decreasing the pulling force.  For very weak forces, however, the entropic effects become significant and hide the effect of topology on the translocation time. Moreover, the simulation time becomes very large and simulations (experiments) are not cost-effective. The results indicate that moderately weak forces can be used to discriminate the three topologies. For this purpose, we calculate the average passage times of the three topologies using a suitable pulling force (considering the bond strength). Then, the maximum and the minimum average passage time is found among the three topologies. Figs. \ref{fig5}(a) and \ref{fig5}(b) show the topologies that have the minimum and the maximum of the average passage times, respectively. We note that changing the dataset used for averaging does not alter the order of the average passage times of the three topologies for all bond strengths and the corresponding suitable forces (Fig. S7 and Table S1). Therefore, the average translocation time can be regarded as a tool for reading the chain topology. (See ESI section 3)

To generalize our findings, we investigate translocation of 5-contact chains to find a correlation between topology of the chain and its average passage time.
In three sets of simulations, one of the cross, parallel, and series relations is taken to be dominant in the topology of the chain, meaning that the majority of the contact pairs have the dominant arrangement. We call such states as ''pure states''. More specifically, 8 out of 10 total binary relations are taken the same; however, the numbers of the other two relations are not determined. 
The average passage time in each set is calculated over 150 randomly chosen chains that fulfill the mentioned conditions; $N_c=8$, $N_p=8$ or $N_s=8$. 
For each bond energy and pulling force, we also calculate the average passage time for a fourth set which contains 150 completely random chains.  

Again, we calculate the average passage times for the three pure states. Then, the maximum and the minimum passage times are found between the three sets. Pure states that have the maximum and the minimum of the average passage times are shown in Figs. \ref{fig5}(c) and \ref{fig5}(d). Extremely large passage times, which occur due to chain knotting, are removed from the data prior to averaging. The order of the average passage time among pure states does not depend on the data set used, while the data set contains enough data points (Table S2). This shows that the dominant topology in pure states can be recognized by using the passage time through a nano-pore. (See ESI section 4)
 
 \section{Conclusions}
In summary, we studied translocation of folded polymers through nano-pores using molecular dynamics simulations. We found settings that are required for a nano-pore setup to be able to read and sort molecules based on their molecular topology. We showed that nano-pores can be used to efficiently enrich certain topologies from mixtures of random 5-contact chains and that this purification is not sensitive to chain orientation in the nano-pore. We also showed that nano-pores can be used to determine the chain topology for 2-contact chains when the intact folded chains pass through the pore. When the chain unfolds upon passing through the nano-pore, we showed that the nano-pore enables determining the position of the contacts along a 2-contact chain in large pulling forces. 
In this condition, by using moderate forces, we could discriminate between pure states (i.e., states for which the majority of contacts were arranged identically) by using the average passage time.

\section*{Acknowledgements}
The authors thank Mahdieh Mikani for technical help and Fatemeh Ramazani for careful reading of the paper.


\end{document}